\documentclass[12pt]{article}
\usepackage{hyperref}
\usepackage[pdftex]{graphicx}
\usepackage{amsmath,amssymb,amsfonts}

\newcommand{\Sc}{Schr\"odinger equation }

\newcommand{\si}{\mbox{\boldmath $\sigma$}}

\newcommand{\q}{\mbox{\boldmath $q$}}

\newcommand{\bb}{\mbox{\boldmath $b$}}

\newcommand{\B}{\mbox{\boldmath $B$}}

\newcommand{\kk}{\mbox{\boldmath $k$}}

\newcommand{\eref}[1]{(\ref{#1})}

\newcommand\fr{\displaystyle\frac}

\newcommand\lt{\left}
\newcommand\rt{\right}

\newcommand\om{\omega}

\begin{document}
\begin{center}
{\Large\bf T-odd correlations in neutron reflectometry
experiments}
\medskip

{\bf V.K.Ignatovich, Yu.V.Nikitenko}
\medskip

{\it Frank Laboratory of Neutron Physics of Joint Institute for
Nuclear Research, 141980, Dubna Moscow region, Russia}
\bigskip

\begin{abstract}
It is shown that transmission amplitudes of magnetic systems with
noncollinear magnetization contain T-odd correlations. Relation of
these T-odd correlations to T-invariance and detailed balance is
discussed.
\end{abstract}
\end{center}

\section{Introduction}
Let's consider reflection and transmission of the magnetic mirror
shown in Fig.~\ref{f1}. The mirror of the total thickness $d$
consists of two magnetic layers. Their magnetizations parallel to
the coordinate plane $x,y$ are  at an angle $\varphi$ to each
other. The outside field is supposed to be zero. We will show that
the neutron transmission matrix amplitude contains a term
proportional to the time odd correlation
\begin{equation}\label{1}
\si\cdot[\B_1\times\B_2],
\end{equation}
where $\si$ is the vector of the Pauli matrices that represents
direction of the neutron spin. Observation of such a correlation
can be interpreted as a violation of the T-invariance.

The problem of T-odd correlations in cross sections of polarized
neutrons in presence of corkscrew like magnetic fields and
fluctuations was first discussed in works [1-4]. We will show that
the T-odd term in the transmission matrix does not mean violation
of the T-invariance even in presence of an absorption, which makes
Hamiltonian to be noninvariant with respect to reverse of time.

In the next section we show how the correlation \eref{1} does
appear in transmission of the system, shown in Fig.~\ref{1}. In
section 3 reflection of the system shown in Fig.~\ref{f1} is
discussed. It is shown that reflection matrix does not contain
T-odd terms, however it violates detailed balance principle.  In
section 4 we discuss T-odd correlation appearing in interaction of
neutrons with a mirror having helicoidal magnetization, and show
that violation of the detailed balance principle here is seen the
most pronounced.

In section 5 we discuss the concept of the T-invariance as applied
to neutron reflectometry, and show that this invariance is not
violated notwithstanding of the appearance of T-odd terms, even if
the system contains an absorption.

\section{Derivation of \eref{1}}

\begin{figure}[t]
{\par\centering\resizebox*{14cm}{!}{\includegraphics{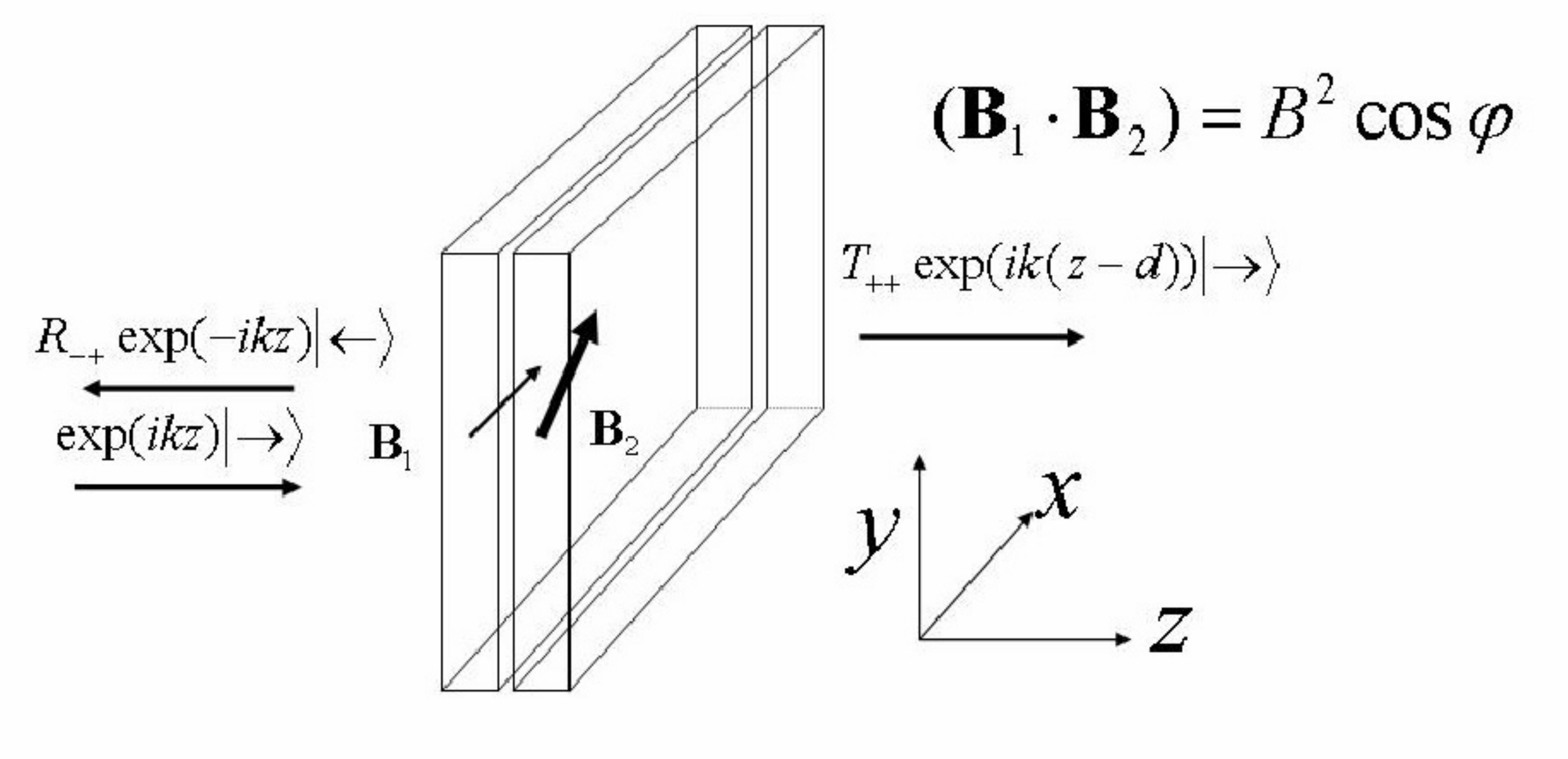}}
\par}
\caption{\label{f1}Reflection and transmission of a magnetic
mirror of thickness $d$, consisting of 2 films magnetized within
$x,y$ coordinate plane, which is parallel to their interfaces.
External magnetic field is zero. Magnetic fields $\B_{1,2}$ of the
films are at an angle $\varphi$ to each other. The incident
neutron going from the left and polarized along the normal to the
films ($z$-axis) can be reflected, say, with spin flip ($R_{-+}$
is the reflection amplitude) or transmitted, say, without spin
flip ($T_{++}$ is the transmission amplitude).}
\end{figure}

Transmission matrix, $T_t$, of the system Fig.~\ref{f1}, if we
neglect multiple reflections between layers\footnote{Inclusion of
multiple reflections does not change the result, but complicates
formulas.}, is represented~\cite{ign,ig0} as
\begin{equation}\label{2}
T_t\approx T_2(\si\B_2)T_1(\si\B_1),
\end{equation}
where $T_{i}$ ($i=1,2$) are transmission matrices of separate
layers:
\begin{equation}\label{3}
T_i(\si\B_i)=\exp(ik'(\si\B_i)l_i)\fr{1-r^2(\si\B_i)}{1-r^2(\si\B_i)\exp(2ik'(\si\B_i)l_i)},
\end{equation}
where $l_i$ is the thickness of the $i$-th layer ($i=1,2$),
\begin{equation}\label{5}
k'(\si\B_i)=\sqrt{k^2-u_j-\si\B_i}
\end{equation}
$k$ is the wave number of the incident neutron, $u_i=£í-iu''$ is
the optical potential of the $i$-th layer, and
\begin{equation}\label{4}
r(\si\B_i)=\fr{k-k'(\si\B_i)}{k+k'(\si\B_i)}
\end{equation}
is the reflection matrix at the interface between vacuum and
$i$-th layer The potential $u$ is defined with the factor
$2m/\hbar^2$, and the field $\B$ is defined with the factor $2\mu
m/\hbar^2$ ($m$ and $\mu$ are the neutron mass and the absolute
value of its magnetic moment respectively).

An arbitrary function $f(\si\B)$ can be represented in the form
\begin{equation}\label{6}
f(\si\B)=f^{(+)}+\si\bb f^{(-)},
\end{equation}
where
\begin{equation}\label{7}
f^{(\pm)}=\fr{f(B)\pm f(-B)}{2},\qquad \bb=\fr{\B}{B}.
\end{equation}
Therefore
\begin{equation}\label{8}
T_i(\si\B_i)=T_i^{(+)}+\si\bb_iT_i^{(-)},
\end{equation}
and its substitution into \eref{2} gives
\begin{equation}\label{9}
T_t=\lt[T_1^{(+)}+\si\bb_1T_1^{(-)}\rt]\lt[T_2^{(+)}+\si\bb_2T_2^{(-)}\rt]=$$
$$=T_1^{(+)}T_2^{(+)}+ T_1^{(-)}T_2^{(+)}\si\cdot\bb_1
+T_2^{(-)}T_1^{(+)}\si\cdot\bb_2+T_1^{(-)}T_2^{(-)}(\si\cdot\bb_1)(\si\cdot\bb_2)=$$
$$=\lt[T_1^{(+)}T_2^{(+)}+T_1^{(-)}T_2^{(-)}\cos\varphi\rt]+ T_1^{(-)}T_2^{(+)}\si\cdot\bb_1
+T_2^{(-)}T_1^{(+)}\si\cdot\bb_2+iT_1^{(-)}T_2^{(-)}(\si\cdot[\bb_1\times\bb_2]),
\end{equation}
where the last term, which contains correlation \eref{1}, results
from the well known relation
\begin{equation}\label{10}
(\si\cdot\bb_1)(\si\cdot\bb_2)=(\bb_1\cdot\bb_2)+i(\si\cdot[\bb_1\times\bb_2]).
\end{equation}

Let's aline $x$-axis along the vector $\B_1$. Then
\begin{equation}\label{11}
\si\cdot\bb_2=\sigma_x\exp(i\varphi\sigma_z),
\end{equation}
and the transmission matrix becomes of the form
\begin{equation}\label{12}
T_t=A+C\sigma_x[1+D\exp(i\varphi\sigma_z)]+iE(\si\cdot[\bb_1\times\bb_2]),
\end{equation}
where $A$, $C$, $D$, $E$ are complex functions of the incident
wave number $k$. From \eref{12} it follows that the transmission
probabilities without spin flip, for initial states
$\sigma_z|\pm\rangle=\pm|\pm\rangle$ are
\begin{equation}\label{13}
T(\pm\pm)=|\langle\pm|T_t|\pm\rangle|^2=|A|^2+|E|^2\pm2{\rm
Im}(AE^*),
\end{equation}
where Im$(x)$ means imaginary part of $x$.

We see that because of the correlation \eref{1} transmissions of
neutrons polarized along and opposite $z$-axis are different. It
can be interpreted as a violation of time-invariance. On the other
hand, if we do not know about the presence of the fields $\B_i$,
we can interpret the difference of non spin-flip transmissions as
a result of correlation $\kk\si$, which violates space
parity-invariance. Such an interpretation looks plausible, because
the term ${\rm Im}(AE^*)$ varies with change of the momentum $k$.
So, we see that interpretation of an experiment is not
unambiguous.

\section{Neutron reflection from the two layer magnetic mirror}

Though transmission of the two layer mirror contains T-odd term,
reflectivity does not contain such terms, however spin-flip
reflection probability violates detailed balance principle.

Reflection matrix amplitude (for simplicity we neglect multiple
scattering) is
\begin{equation}\label{r1}
R_t\approx R_1(\si\B_1)+T_1(\si\B_1)R_2(\si\B_2)T_1(\si\B_1),
\end{equation}
where
\begin{equation}\label{r2}
R(\si\B_i)=r(\si\B_i)\fr{1-\exp(2ik'(\si\B_i)l_i)}{1-r^2(\si\B_i)\exp(2ik'(\si\B_i)l_i)}.
\end{equation}
\begin{figure}[t]
{\par\centering\resizebox*{16cm}{!}{\includegraphics{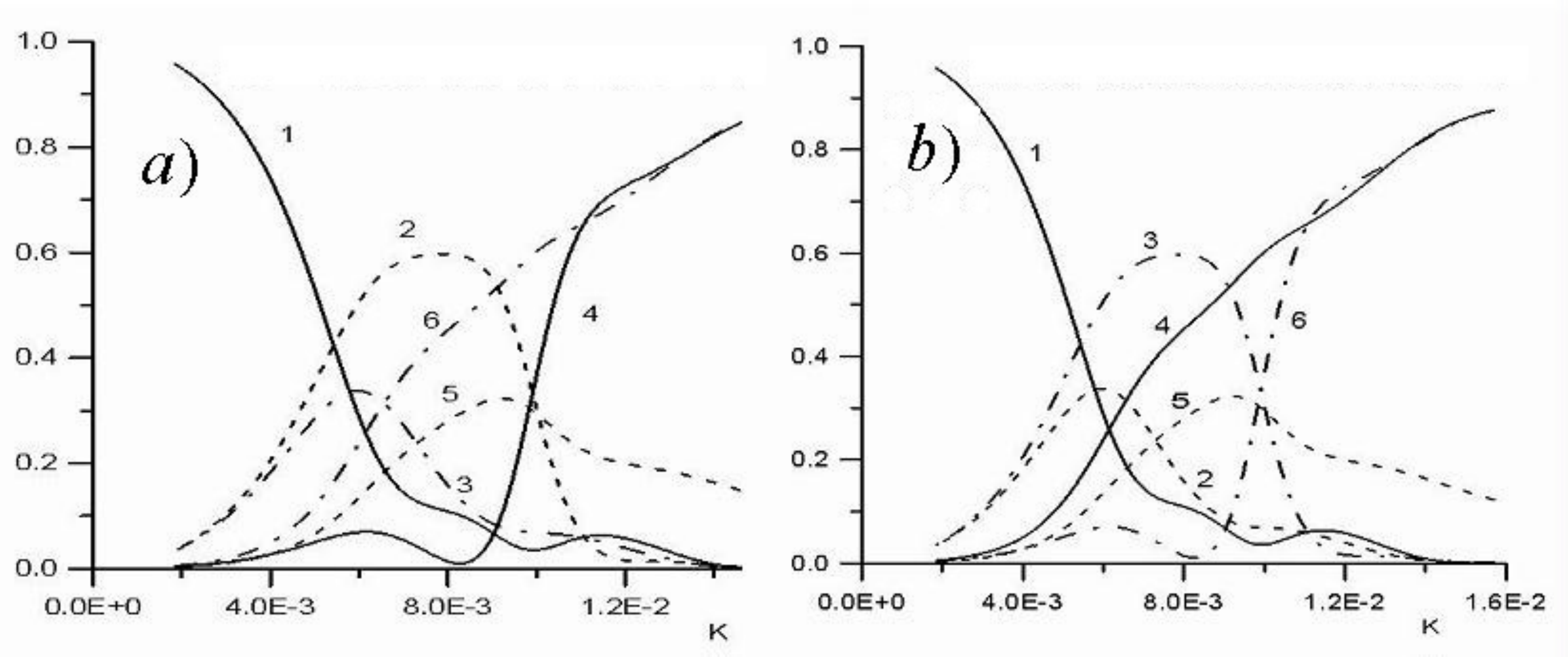}}
\par}
\caption{\label{f1a}Calculated reflection, $R(ij)=|R_{ij}|^2$, and
transmission, $T(ij)=|T_{ij}|^2$, ($i,j=\pm$) coefficients of the
two layer mirror of the same thickness when angle $\varphi$
between magnetizations is a) $\pi/2$, and b) $-\pi/2$. The curves correspond to
1 --- $R(+ +)=R(- -)$, 2 --- $R(-+)$, 3 --- $R(+-)$, 4 --- $T(++)$, 5 --- $T(+-)=T(-+)$, 6 --- $T(- -)$.
The change of sign of $\varphi$ leads to exchange
$R(\pm\mp)\to R(\mp\pm)$ of spin-flip curves in reflectivities and
of non spin-flip $T(\om\pm)\to T(\mp\mp)$ curves in
transmissivities.}
\end{figure}

Substitution of \eref{r2} and \eref{3} into \eref{r1} with account
of the representation \eref{6} and choice of $x$-axis along vector
$\B_1$ reduces \eref{r1} to the form
\begin{equation}\label{r3}
R_t=A+C\sigma_x[1+D\exp(i\varphi\sigma_z)+E\exp(-i\varphi\sigma_z)],
\end{equation}
where $A$, $C$, $D$, $E$ are complex functions of the incident
wave number $k$. From this expression we see that non spin-flip
reflectivities for both incident polarizations are equal to each
others:
\begin{equation}\label{r4}
R(++)=R(--)=|\langle\pm|R_t|\pm\rangle|^2,
\end{equation}
while spin-flip reflectivities, $R(\pm\mp)$, are different and
their dependence on the angle $\varphi$ is such that
\begin{equation}\label{r5}
R(\pm\mp,\varphi)=R(\mp\pm,-\varphi),
\end{equation}
which is clearly seen in Fig~\ref{f1a}, where reflectivities and
transmissivities calculated for two identical layers of thickness
25 nm, when angle between their magnetizations is
$\varphi=\pm\pi/2$, are shown.

We see that non spin transmissivities are different and change
with sign of $\varphi$ according to the T-odd correlation
\eref{1}. The spin-flip transmissivities are equal because
according to \eref{9} the factor $D$ in \eref{12} is equal to
unity.

Difference of two spin-flip reflectivities means violation of the
detailed balance principle, because it creates a cycle current in
phase space, which diminishes the entropy. We will discuss this
effect in the next section, where violation of the detailed
balance is seen more strikingly.

\section{Neutron reflectometry for a magnetic mirror with helicoidal magnetization}

\begin{figure}[t!]
{\par\centering\resizebox*{14cm}{!}{\includegraphics{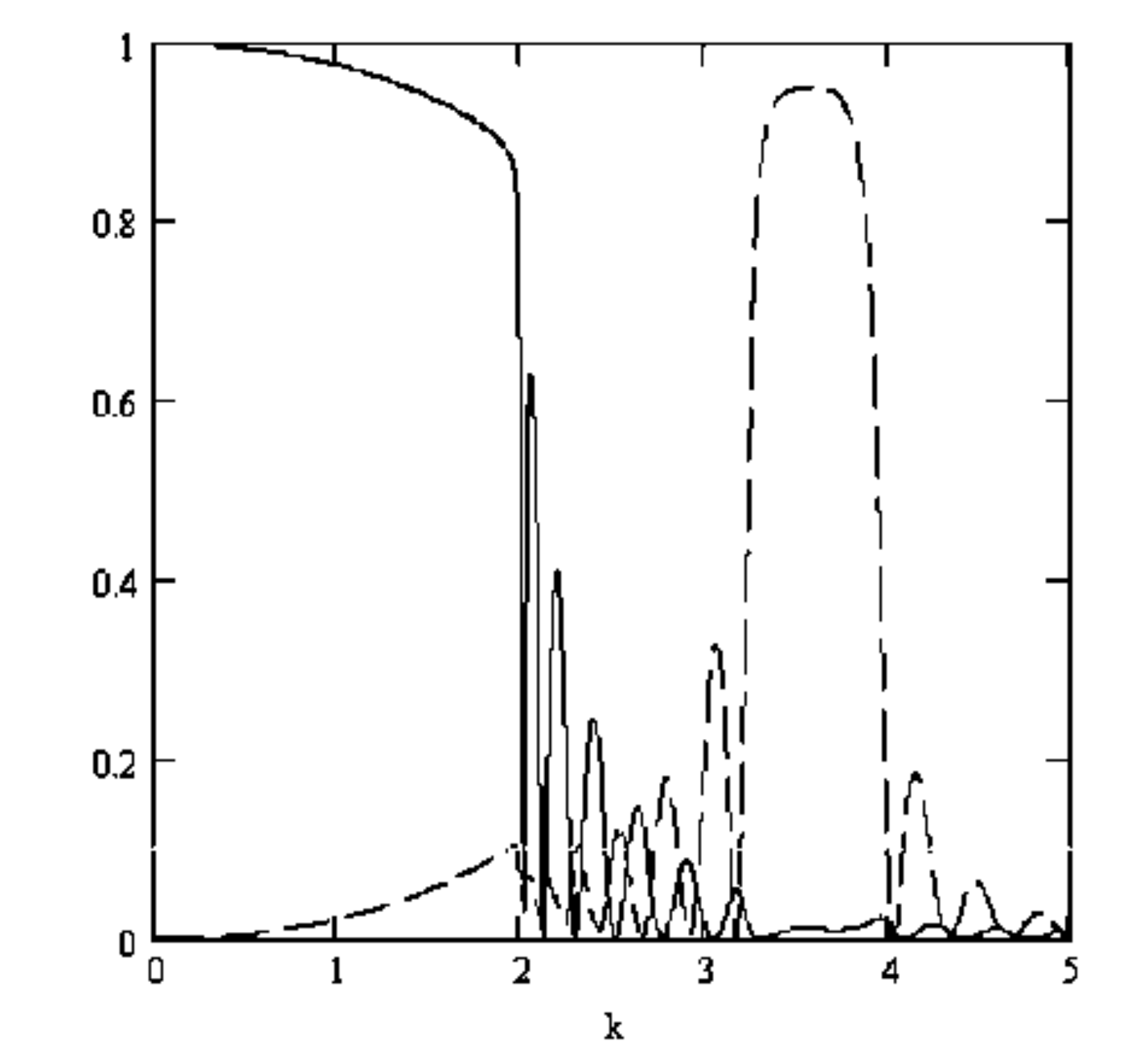}\hfill\includegraphics{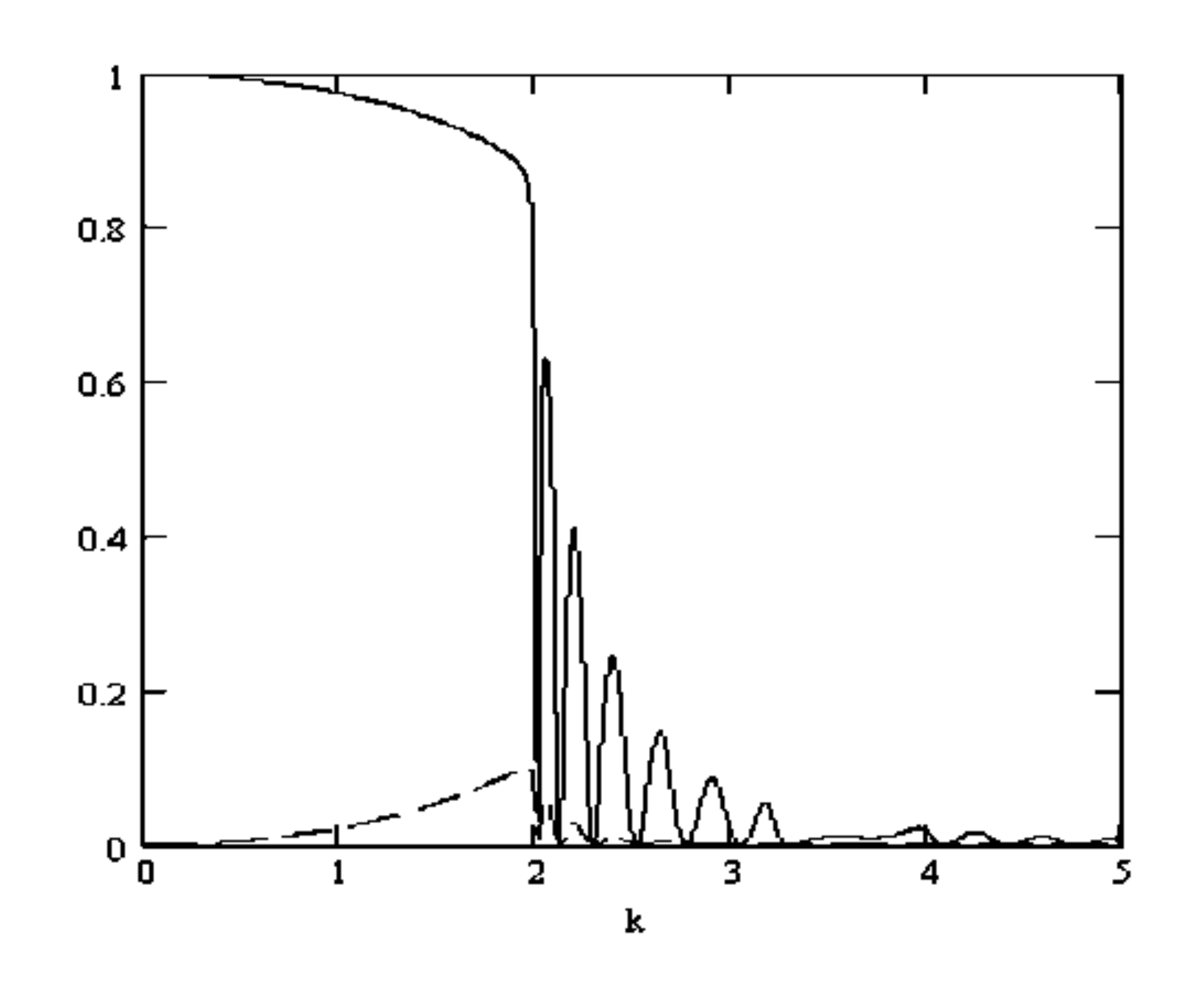}}
\par}
\caption{\label{f2}Reflectivity of a helicoidal magnetic mirror
with (dashed line) and without spin flip (solid line) for the
incident polarization opposite (left) and along (right)
$z$-axis~\cite{len}. We see that above the range of total
reflection there is a well pronounced peak of almost total
reflection with spin flip, when the incident neutron is polarized
against $z$-axis.}
\end{figure}

\begin{figure}[b!]
{\par\centering\resizebox*{14cm}{!}{\includegraphics{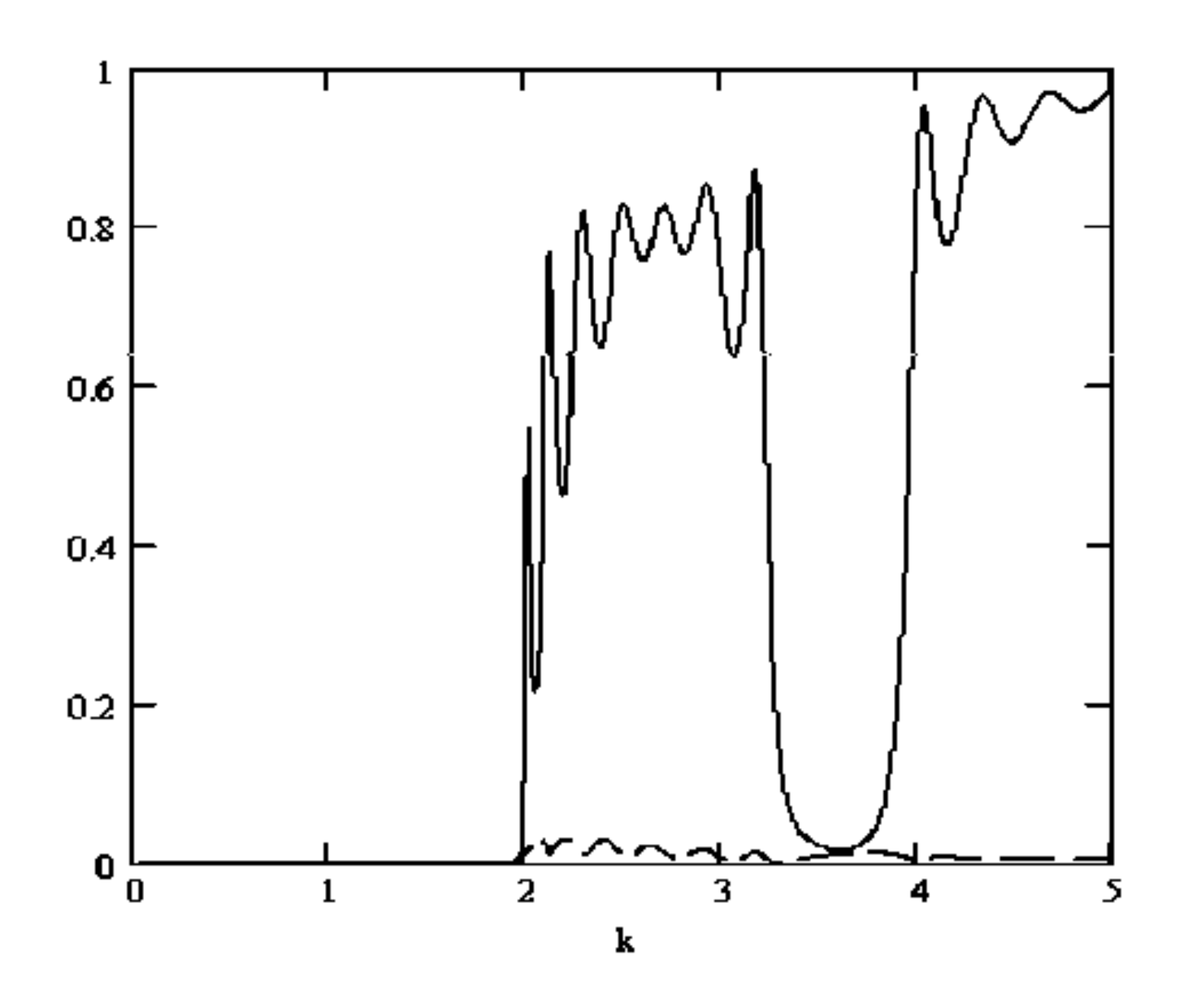}\hfill\includegraphics{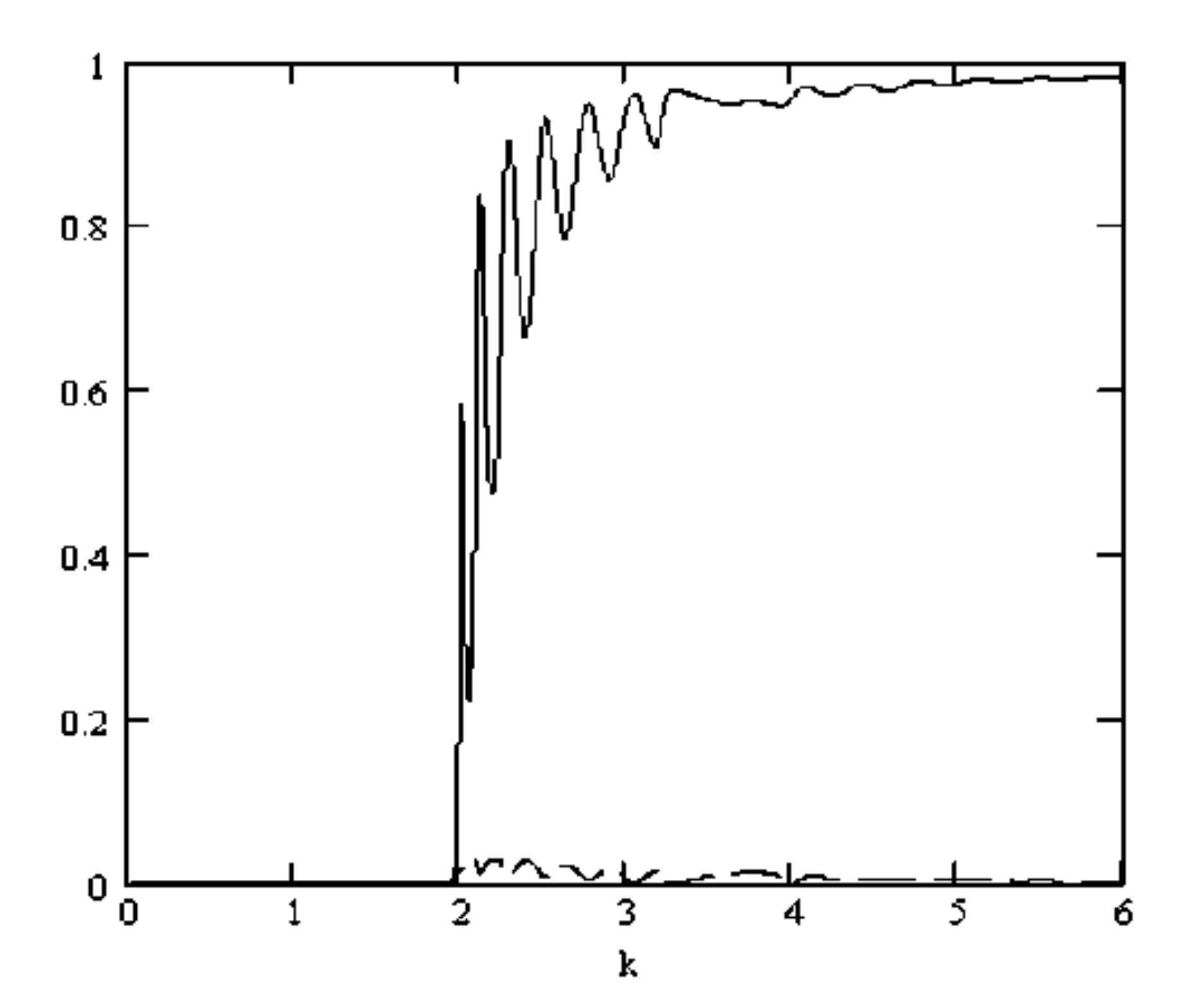}}
\par}
\caption{\label{f3}Transmission probabilities of a helicoidal
magnetic mirror with (dashed line) and without spin flip (solid
line) for the incident polarization opposite (left) and along
(right) $z$-axis~\cite{len}. We see that above the range of total
reflection there is a well pronounced dip in transmission of
neutrons initially polarized against $z$-axis.}
\end{figure}

\begin{figure}[h!]
{\par\centering\resizebox*{12cm}{!}{\includegraphics{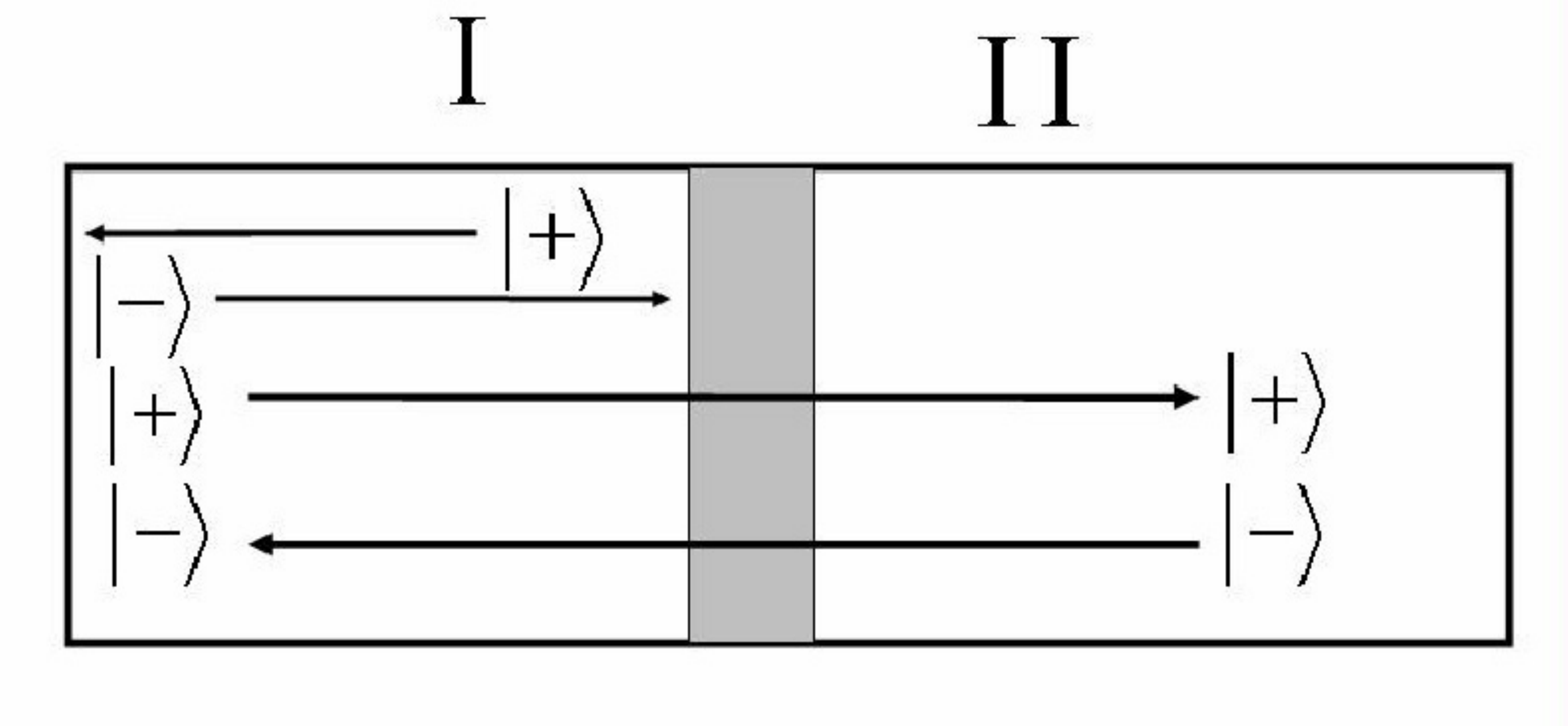}}
\par}
\caption{\label{f4}Illustration of violation of the detailed
balance principle.}
\end{figure}

The false effect of the time and parity violation is especially
well seen in the case of the neutron reflection from a magnetic
mirror magnetized  helicoidally~\cite{len} around a vector $\q$
which is directed along $z$-axis parallel to the normal to the
mirror interface. Neutron wave function in helicoidal field was
found in~\cite{calv}, though reflection and transmission of
helicoidal mirrors were calculated in~\cite{fra,len}. In
Fig.~\ref{f2} there are shown reflectivities with and without spin
flip for polarizations of the incident neutron along and opposite
$z$-axis. Outside of the mirror magnetic field is absent. The
analogous transmission probabilities are shown in Fig.~\ref{f3}.
We see that there is again a time-odd correlation $\si\q$, which
can be also interpreted as a parity odd correlation with the
incident neutron momentum $\si\kk$.
The resonant spin-flip reflectivity for $|-\rangle$ polarization
increases with the mirror thickness, and becomes almost total.
Such a reflectivity violates the detailed balance principle, and
the violation in this example is especially well seen. Indeed,
imagine that a vessel with ideal walls is homogeneously filled
with a gas of unpolarized neutrons. If we split the vessel into
two parts as shown in Fig.~\ref{f4}, inserting the helicoidal
mirror, then
all the neutrons from the left part I go through the mirror to the
right part II, and become completely polarized along $z$-axis.
Indeed, the neutrons in state $|+\rangle$ go directly through the
mirror, and cannot go back, while the neutrons in the state
$|-\rangle$ are reflected from the mirror with spin-flip, and
after reflection from ideal walls of the vessel go again to the
mirror and through it to the part II. Therefore it looks as if all
the neutrons from the part I gather in the part II in the single
state $|+\rangle$, which terribly decreases the entropy.

However in fact such a spilling over the mirror from left to right
is compensated by the opposite flux from II to I, because in the
right part neutrons in the state $|-\rangle$ can go through the
mirror, while neutrons in the state $|+\rangle$ are reflected from
the mirror with spin flip to the state $|-\rangle$ and at the
second incidence go through the mirror to the part I. Therefore in
both parts neutrons remain isotropic and in unpolarized state,
however in the phase space there appear a cycle, which means a
decrease of the entropy.

Decrease of the entropy is created by the rotation vector $\q$.
Such a rotation makes the space into imbalance. Even two magnetic
fields considered in previous section create an imbalance, though
this imbalance is not so evident as in the considered case of the
helicoidal mirror.

\section{Analysis of the time invariance}

Let's analyze the principle of T-invariance, in the simplest case
of the neutron scattering on a nonmagnetic one-dimensional
potential $u(x)$, which is nonzero in an interval $0\le x\le d$.
The neutron wave function outside of the potential is
\begin{equation}\label{1a3}
\psi(x,t)=e^{-i\omega
t}\lt[\Theta(x<0)\Big(\exp(ikx)+R(k)\exp(-ikx)\Big)+\Theta(x>d)T(k)\exp(ik(x-d))
\rt],
\end{equation}
where $\Theta(x)$ is a step function equal to unity when
inequality in its argument is satisfied, and to zero otherwise,
and $R(k)$ and $T(k)$ are reflection and transmission amplitudes,
which are complex function of the incident wave number $k$.  The
wave function is a solution of the \Sc
\begin{equation}\label{14}
\lt(i\fr{\partial}{\partial t}+\fr{\partial^2}{\partial
x^2}-u(x)\rt)\psi(x,t)=0.
\end{equation}
If we make a transformation
\begin{equation}\label{16}
t\to-t
\end{equation}
the equation for $\psi(x,-t)$ will look differently comparing to
\eref{14}. To restore its form we have to make complex
conjugation, after which we get
\begin{equation}\label{15}
\lt(i\fr{\partial}{\partial t}+\fr{\partial^2}{\partial
x^2}-u^*(x)\rt)\psi^*(x,-t)=0.
\end{equation}
However we are to be careful here. The potential, if it has an
imaginary part, changes after complex conjugation, therefore we
cannot be sure that the function $\psi^*(x,-t)$ remains a solution
of \eref{15}. Therefore, instead of \eref{15} we should write
\begin{equation}\label{15a}
\lt(i\fr{\partial}{\partial t}+\fr{\partial^2}{\partial
x^2}-u^*(x)\rt)\Psi(x,t)=0,
\end{equation}
and check, whether $\Psi(x,t)=\psi^*(x,-t)$ or not. We will prove
that this equality is true in the case of a rectangular potential,
and claim that there are no reason to doubt its truth for other
potentials.

In fact we can deal with stationary equations, representing
$\Psi(x,t)=\exp(-i\omega t)\Phi(x)$ and $\psi(x,t)=\exp(-i\omega
t)\phi(x)$, and our goal is to show that solution $\Phi(x)$ of the
equation
\begin{equation}\label{15a1}
\lt(k^2+\fr{\partial^2}{\partial x^2}-u^*(x)\rt)\Phi(x)=0,
\end{equation}
coincides with $\phi^*(x)$.

In the case of a rectangular barrier potential of height
$u=u'-iu''$ and width $d$ the wave function on the full axis $x$
is $\phi(x,u)$, where~\cite{lan}
\begin{equation}\label{13a}
\phi(x,u)=\Theta(x<0)\Big(e^{ikx}+R(k,u)e^{-ikx}\Big)+$$
$$+\Theta(0<x<d)\fr{[1+r(k,u)]\exp(ik'(u)d)}{1-r^2(k,u)\exp(2ik'(u)d)}\lt[e^{ik'(u)(x-d)}-r(k,u)e^{-ik'(u)(x-d)}\rt]
+$$ $$+\Theta(x>d)T(k,u)e^{ik(x-d)},
\end{equation}
where
\begin{equation}\label{1a}
R(k,u)=\fr{r(k,u)[1-\exp(2ik'(u)d)]}{1-r^2(k,u)\exp(2ik'(u)d)},
\qquad
T(k,u)=\fr{\exp(ik'(u)d)[1-r^2(k,u)]}{1-r^2(k,u)\exp(2ik'(u)d)},
\end{equation}
\begin{equation}\label{4a}
r(k,u)=\fr{k-k'(u)}{k+k'(u)},\qquad k'(u)=\sqrt{k^2-u},
\end{equation}
and everywhere we explicitly pointed out dependence on the complex
potential $u$. The function $\phi^*(x,u)$ is
\begin{equation}\label{3a}
\phi^*(x,u)=\Theta(x<0)\Big(e^{-ikx}+R^*(k,u)e^{ikx}\Big)+$$
$$+\Theta(0<x<d)\fr{[1+r^*(k,u)]\exp(-ik'^*(u)d)}{1-r^{*2}(k,u)\exp(-2ik'^*(u)d)}
\lt[e^{-ik'^*(u)(x-d)}-r^*(k,u)e^{+ik'^*(u)(x-d)}\rt] +$$
$$+\Theta(x>d)T^*(k,u)e^{-ik(x-d)}.
\end{equation}
The function \eref{3a} describes interference of two waves
incident from the left and right with relative amplitudes $R^*(k)$
and $T^*(k)$. We will show that it coincides with the solution
$\Phi(x)$ of \eref{15a1} containing these two incident waves. The
wave incident from the left gives a solution $\Phi_l(x)$, the
wave, incident from the right gives a solution $\Phi_r(x)$, and
the total solution is $\Phi_l(x)+\Phi_r(x)$. With the help of
general approach presented in~\cite{lan,igna} we obtain
\begin{equation}\label{2a}
\Phi_l(x,u^*)=\Theta(x<0)R^*(k,u)\Big(e^{ikx}+R(k,u^*)e^{-ikx}\Big)+$$
$$+\Theta(0<x<d)R^*(k,u)\fr{[1+r(k,u^*)]\exp(ik'(u^*)d)}{1-r^2(k,u^*)\exp(2ik'(u^*)d)}
\lt[e^{ik'(u^*)(x-d)}-r(k,u^*)e^{-ik'(u^*)(x-d)}\rt] +$$
$$+\Theta(x>d)R^*(k,u)T(k,u^*)e^{ik(x-d)}
\end{equation}
is the wave function for the incident wave $R^*(k,u)\exp(ikx)$,
and
\begin{equation}\label{4a1}
\Phi_r(x,u^*)=\Theta(x<0)T^*(k,u)T(k,u^*)e^{-ikx}+$$
$$+\Theta(0<x<d)T^*(k,u)\fr{[1+r(k,u^*)]\exp(ik'(u^*)d)}{1-r^2(k,u^*)\exp(2ik'(u^*)d)}
\lt[e^{-ik'(u^*)x}-r(k,u^*)e^{ik'(u^*)x}\rt]
+$$
$$+\Theta(x>d)T^*(k,u)\lt(T(k,u^*)e^{-ik(x-d)}+R(k,u^*)e^{ik(x-d)}\rt)
\end{equation}
is the wave function for the incident wave $T^*(k)e^{-ik(x-d)}$.
From \eref{4a} and \eref{1a} it is seen, that $k'(u^*)=k'^*(u)$,
and $r(u^*)=r^*(u)$, but $R(k,u^*)\ne R^*(k,u)$, and $T(k,u^*)\ne
T^*(k,u)$.

It is easy to verify by by simple algebra that the sum of terms in
the interval $0<x<d$ from \eref{2a} and \eref{4a1} is equal to the
middle term in \eref{3a}. This algebra is shown symbolically in
the following 6 lines where $K'$ denotes $k'(u^*)$, and in other
terms dependence on $k$ and $u$ is omitted:
\begin{equation}\label{5a1}
R^*\fr{[1+r^*]e^{iK'd}}{1-r^{*2}e^{2iK'd}}\lt[e^{iK'(x-d)}-r^*e^{-iK'(x-d)}\rt]
+T^*\fr{[1+r^*]e^{iK'd}}{1-r^{*2}e^{2iK'd}}\lt[e^{-iK'x}-re^{iK'x}\rt]
=$$
$$=\fr{r^*(1-e^{-2ik'^*d})}{1-r^{*2}e^{-2ik'^*d}}\fr{[1+r^*]e^{iK'd}}{1-r^{*2}e^{2iK'd}}
\lt[e^{iK'(x-d)}-r^*e^{-iK'(x-d)}\rt]+$$
$$+\fr{e^{-ik'^*d}(1-r^{*2})}{1-r^{*2}e^{-2ik'^*d}}\fr{[1+r^*]}{1-r^{*2}e^{2iK'd}}
\lt[e^{-iK'(x-d)}-e^{2iK'd}r^*e^{iK'(x-d)}\rt]=$$
$$e^{ik'^*(x-d)}\Big[r^*(1-e^{-2ik'^*d})[1+r^*]e^{ik'^*d}-e^{-ik'^*d}(1-r^{*2})[1+r^*]e^{2ik'^*d}r^*\Big]+$$
$$+e^{-ik'^*(x-d)}\Big[-r^*(1-e^{-2ik'^*d})[1+r^*]e^{ik'^*d}r^*+e^{-ik'^*d}(1-r^{*2})[1+r^*]\Big]=$$
$$=\fr{[1+r^*]e^{-ik'^*d}}{1-r^{*2}e^{-2ik'^*d}}\Big[e^{-ik'^*(x-d)}-r^*e^{ik'^*(x-d)}\Big]\end{equation}
In a similar way it is possible to verify that the sum of
amplitudes of two outgoing waves at $x<0$ is equal to
\begin{equation}\label{18}
R^*(k,u)R(k,u^*)+T^*(k,u)T(k,u^*)=1.
\end{equation}
The right outgoing wave at $x>d$ vanishes. Its amplitude
\begin{equation}\label{19}
R^*(k,u)T(k,u^*)+T^*(k,u)R(k,u^*)=2{\rm Re}(R^*(k,u)T(k,u^*))=0,
\end{equation}
which shows that the phases of the amplitudes $R(k,u)$ and
$T(k,u^*)$ differ by $\pi/2$. So, we see that the scattering of a
scalar particle on a complex potential is time reversible. We have
checked it for a simple rectangular potential,  but there are no
reason to think that for more complex potential the result will be
different.

In a similar way it is possible to show that scattering of a
spinor particle on an arbitrary magnetic potential will be
reversible after transformation of \eref{16}, reverse of fields
and spin and complex conjugation. The transformed wave function
will describe the time reversed processes.

\section{Conclusion}
With the help of simple examples we have shown how in simple
neutron reflectometry experiment can appear T-odd terms that can
be interpreted as T- or P-parity violation though they do not
violate nor T- nor P-invariance. At the same time, if the space
contains even a couple of noncollinear magnetic fields,
interaction of neutrons with this couple does not satisfy the
principle of detailed balance.

\section*{Acknowledgment}
This work is supported by RFBR grant 08-02-00467a.
The authors are grateful to A.A.Fraerman for his suggestions,
and also to G. Petrov, Yu. Chuvilskiy and
V.Bunakov for their interest and discussion.

\end{document}